\documentclass[conference,comsoc]{IEEEtran}

\usepackage[T1]{fontenc}
\usepackage{cite}
\usepackage{amsmath}
\usepackage{amssymb}
\usepackage{mathtools}
\usepackage{siunitx}
\DeclareSIUnit\dBm{dBm}
\usepackage{bm}
\usepackage{commath}
\usepackage{enumitem}
\usepackage{adjustbox}
\usepackage{array}
\usepackage{nth}

\usepackage[cmintegrals]{newtxmath}
\usepackage[caption=false,font=footnotesize]{subfig}
\usepackage{url}
\usepackage{marginnote}
\usepackage{blindtext}

\DeclareMathOperator{\jj}{j}

\let\originalleft\left
\let\originalright\right
\renewcommand{\left}{\mathopen{}\mathclose\bgroup\originalleft}
\renewcommand{\right}{\aftergroup\egroup\originalright}

\usepackage{tikz}
\usepackage{pgfplots}
\usepackage{pgfplotstable}
\pgfplotsset{compat=newest}
\usepgfplotslibrary{colormaps, groupplots, patchplots}
\usetikzlibrary{external}
\usetikzlibrary{arrows.meta, backgrounds, calc, decorations, decorations.markings, decorations.pathreplacing, math, patterns, pgfplots.polar, shapes}
\pgfdeclarelayer{bg}    
\pgfsetlayers{bg,main}  
\usepackage{tkz-euclide}

\usepackage[acronym, automake]{glossaries-prefix}
\setacronymstyle{long-short}
\newacronym{1D}{1D}{one-dimensional}
\newacronym{2D}{2D}{two-dimensional}
\newacronym{3D}{3D}{three-dimensional}
\newacronym{3GPP}{3GPP}{Third Generation Partnership Project}
\newacronym{5G}{5G}{fifth generation}
\newacronym{6G}{6G}{sixth generation}
\newacronym{ADC}{ADC}{analog-digital converter}
\newacronym{ADCGM}{ADCGM}{alternating descent conditional gradient method}
\newacronym{AFDM}{AFDM}{affine frequency-division multiplexing}
\newacronym[longplural=angles of arrival]{AOA}{AOA}{angle of arrival}
\newacronym[longplural=angles of departure]{AOD}{AOD}{angle of departure}
\newacronym{AWGN}{AWGN}{additive white Gaussian noise}
\newacronym{BER}{BER}{bit error rate}
\newacronym{BPSK}{BPSK}{binary phase-shift keying}
\newacronym{BS}{BS}{base station}
\newacronym{cdf}{cdf}{cumulative distribution function}
\newacronym{CP}{CP}{cyclic prefix}
\newacronym{CRLB}{CRLB}{Cram\'er-Rao lower bound}
\newacronym{CPM-LFM}{CPM-LFM}{continuous phase modulated-linear frequency modulated}
\newacronym{DFnT}{DFnT}{discrete Fresnel transform}
\newacronym{DFT}{DFT}{discrete Fourier transform}
\newacronym{DL}{DL}{downlink}
\newacronym{EFIM}{EFIM}{equivalent Fisher information matrix}
\newacronym{ESPEB}{ESPEB}{expected \glsentrytext{SPEB}}
\newacronym{EV}{EV}{eigenvalue}
\newacronym{EXIP}{EXIP}{extended invariance principle}
\newacronym[longplural=Fisher information matrices]{FIM}{FIM}{Fisher information matrix}
\newacronym[prefixfirst={a\ }, prefix={an\ }]{FMCW}{FMCW}{frequency-modulated continuous wave}
\newacronym{IDFT}{IDFT}{inverse DFT}
\newacronym{iid}{iid}{independent and identically distributed}
\newacronym{IIoT}{IIoT}{industrial internet of things}
\newacronym{ISAC}{ISAC}{integrated sensing and communications}
\newacronym{LOS}{LOS}{line-of-sight}
\newacronym[prefixfirst={a\ }, prefix={an\ }]{MMSE}{MMSE}{minimum mean square error}
\newacronym{MIMO}{MIMO}{multiple-input multiple output}
\newacronym{mm-Wave}{mm-Wave}{millimeter-wave}
\newacronym{MT}{MT}{mobile terminal}
\newacronym{NLOS}{NLOS}{non-LOS}
\newacronym{NR}{NR}{new radio}
\newacronym{OCDM}{OCDM}{orthogonal chirp-division multiplexing}
\newacronym{OFDM}{OFDM}{orthogonal frequency-division multiplexing}
\newacronym{OTDOA}{OTDOA}{observed time difference of arrival}
\newacronym{OTFS}{OTFS}{orthogonal time frequency space}
\newacronym{PAPR}{PAPR}{peak-to-average power ratio}
\newacronym{pdf}{pdf}{probability density function}
\newacronym{PC-FMCW}{PC-FMCW}{phase coded frequency-modulated continuous-wave}
\newacronym{PEB}{PEB}{position error bound}
\newacronym{PMCW}{PMCW}{phase-modulated continuous wave}
\newacronym{pmf}{pmf}{probability mass function}
\newacronym{PSD}{PSD}{positive semidefiniteness}
\newacronym{RCRLB}{RCRLB}{root \glsentrytext{CRLB}}
\newacronym{RE}{RE}{resource element}
\newacronym{RF}{RF}{radio frequency}
\newacronym{RMSE}{RMSE}{root mean square error}
\newacronym{RTT}{RTT}{round-trip time}
\newacronym[prefixfirst={a\ }, prefix={an\ }]{Rx}{Rx}{receiver}
\newacronym[prefixfirst={a\ }, prefix={an\ }]{SDP}{SDP}{semidefinite program}
\newacronym{SFFT}{SFFT}{symplectic finite Fourier transform}
\newacronym{SL}{SL}{sidelink}
\newacronym{SPEB}{SPEB}{squared position error bound}
\newacronym{SNR}{SNR}{signal-to-noise ratio}
\newacronym[longplural=times of arrival]{TOA}{TOA}{time of arrival}
\newacronym{Tx}{Tx}{transmitter}
\newacronym{UE}{UE}{user equipment}
\newacronym{ULA}{ULA}{uniform linear array}
\newacronym{UCA}{UCA}{uniform circular array}
\newacronym{UL}{UL}{uplink}
\newacronym{UTDOA}{UTDOA}{uplink TDOA}
\newacronym{VA}{VA}{virtual anchor}
\newacronym{VCO}{VCO}{voltage-controlled oscillator}
\makeglossaries

\newsavebox\glsscratchboxa
\newsavebox\glsscratchboxb
\newsavebox\glsscratchboxc
\newsavebox\glsscratchboxd

\usepackage{graphicx}

\title{Towards Integrated Sensing and Communications for 6G}

\author{\IEEEauthorblockN{Qi~Wang\IEEEauthorrefmark{1}, Anastasios~Kakkavas\IEEEauthorrefmark{1}\IEEEauthorrefmark{2}, Xitao~Gong\IEEEauthorrefmark{1} and Richard A. Stirling-Gallacher\IEEEauthorrefmark{1} }
	\IEEEauthorblockA{\IEEEauthorrefmark{1}Munich~Research~Center,~Huawei~Technologies~Duesseldorf~GmbH,~Munich,~Germany}
	\IEEEauthorblockA{\IEEEauthorrefmark{2}Department~of~Electrical~and~Computer Engineering, Technische Universit{\"{a}}t M{\"{u}}nchen, Munich, Germany}
	\IEEEauthorblockA{
		\texttt{\{qi.wang1, anastasios.kakkavas, xitao.gong, richard.sg\}@huawei.com}
	}
}

\begin{document}
	
	\bstctlcite{MyBSTcontrol}
	
	
	\maketitle
	\tikzexternaldisable
	\tikz[overlay,remember picture]
	{
		\node at ($(current page.south west)+(0.6in,0.3cm)$) [rotate=0, anchor=south west] {\parbox{\textwidth}{\footnotesize \footnotesize \textcopyright 2021 IEEE. Personal use of this material is permitted. Permission from IEEE must be obtained for all other uses, in any current or future media, including reprinting/republishing this material for advertising or promotional purposes, creating new collective works, for resale or redistribution to servers or lists, or reuse of any copyrighted component of this work in other works.}};
	}
	\tikzexternalenable	
	\begin{abstract}
		For the next generation of mobile communications systems, the integration of sensing and communications promises benefits in terms of spectrum utilization, cost, latency, area and weight. In this paper, we categorize and summarize the key features and technical considerations for different integration approaches and discuss related waveform design issues for a future \gls{6G} system. We provide results on new candidate waveforms for monostatic sensing and finally highlight important open issues and directions that deserve future in-depth research.
	\end{abstract}
	
	
	\IEEEpeerreviewmaketitle

	\glsresetall
	\vspace{-0.2cm}
	\section{Introduction}
		\label{sec:introduction}
		
		Traditionally, sensing and wireless communications have been performed individually using separated functional entities and frequency bands. For the \gls{6G} of mobile communications systems, the goal is to integrate sensing and communications together, so that services can be provided based on both sensing and communications functionalities. This is further motivated by the intended usage of the higher frequency bands (beyond 70 GHz) for \gls{6G}, which aims to facilitate the provisioning of sensing functionality with high-resolution capabilities.
		
		The topic of \gls{ISAC} has caught researchers' interests for at least last ten years~\cite{SW11}. As \gls{6G} research fully accelerates, given various global initiatives, national funded projects and academic/industrial research programs, this topic has become increasingly important and received renewed attention recently \cite{LCM+21, CLJ+21}.  
		
		The aim of this paper is to provide an overview of the key features and technical considerations of different integration approaches for \gls{6G} systems. Unlike the existing surveys, we mainly focus on the related waveform design considerations for fully integrated systems. Furthermore, initial results on a promising chirp-based waveform are presented and discussed.
		
		The paper is structured as follows: Section II describes typical sensing and communications scenarios, followed by a discussion of different integration approaches and waveform considerations in Section III. Section IV describes open issues that deserve further in-depth research and finally in Section V the conclusions are presented.

	\section{Scenarios}
	\label{sec:scenarios}
	
		In this section, we first review typical operating scenarios for sensing systems and for mobile communication systems. Subsequently, potential operating scenarios for an integrated system are presented.
		
		\subsection{Typical Sensing and Communications Scenarios}
			Sensing refers to the technology of retrieving information about an object from measurements of the emitted or reflected signals. The exemplary information of interest are distance/range, position, shape and velocity etc. Well-known sensing applications include automotive radar enabling adaptive cruise control and collision avoidance, consumer devices recognizing hand gestures and body/baggage scanning systems at airport terminals. If the object is equipped with its own \gls{Rx} and/or \gls{Tx} for the sensing signal, the sensing system is considered to be active; otherwise it is known as passive. From a transceiver deployment point of view, a sensing system can be described as monostatic or bi-static. For the former, the \gls{Tx} and the \gls{Rx} of the radio signal are identical or co-located; while for the latter, the \gls{Tx} and the \gls{Rx} are at different locations. 
			
			The most widely used communications system is the mobile radio communications system, which consists of networked radio access nodes, i.e. base stations, and \glspl{UE}. Three types of radio links are specified, namely \gls{DL} for transmission from a base station to a \gls{UE}, \gls{UL} for transmission from a \gls{UE} to a base station and \gls{SL} for transmission between two \glspl{UE}. One of the key features that distinguishes current mobile radio systems from sensing systems is the centralized scheduling and coordination of the radio resource. Given the scarcity of radio spectrum, such a mechanism guarantees the performance, reliability and usability of communications services.
			
		\subsection{Scenarios for an \gls{ISAC} system}
\begin{figure*}
				\centering
				\includegraphics{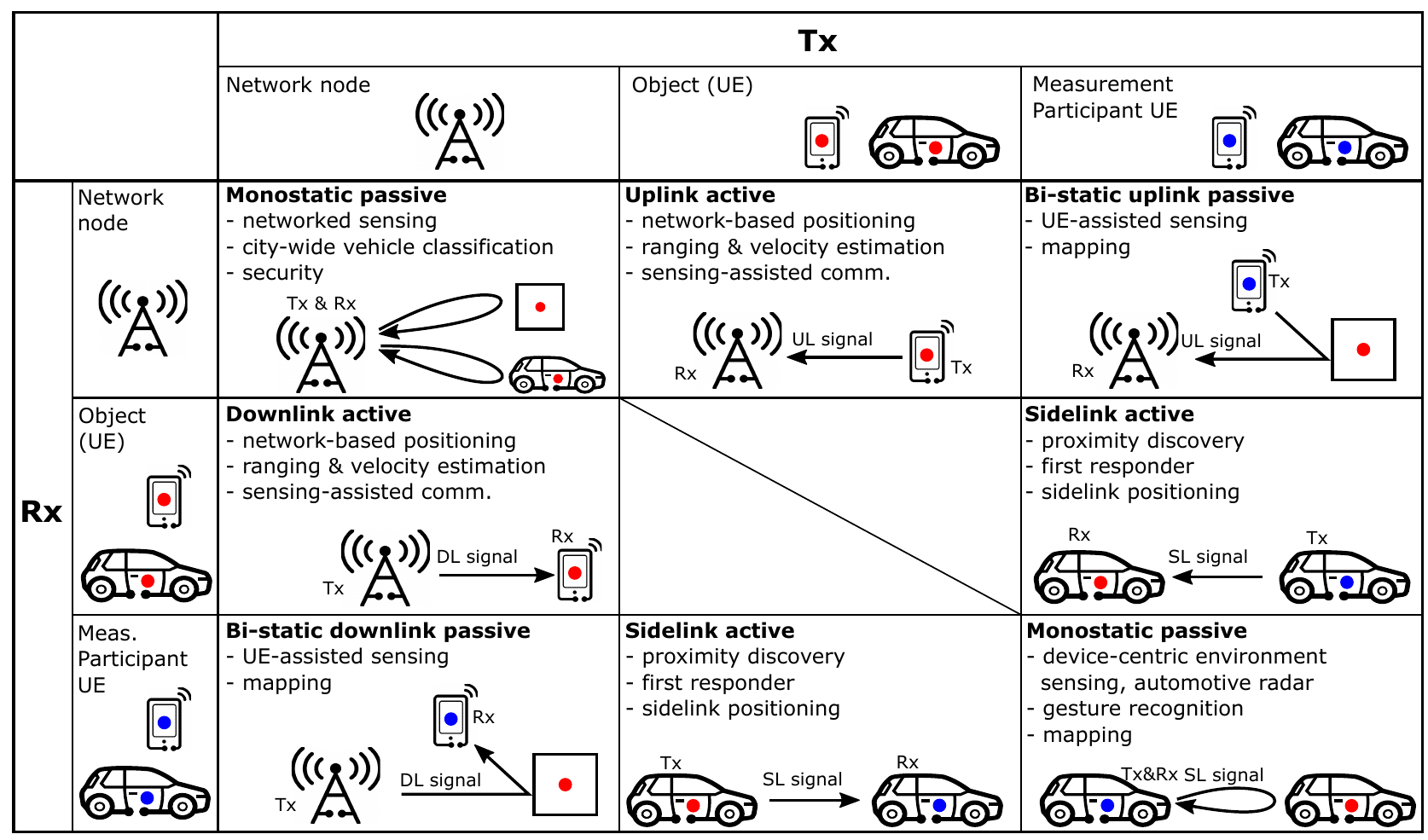}
				\sbox\glsscratchboxa{\footnotesize \gls{ISAC}}
				\caption{An overview of exemplary \unhcopy\glsscratchboxa~operating scenarios and their applications.}
				\vspace{-0.3cm}
				\label{fig:ISAC_operating_scenarios}
			\end{figure*}

			A radio link can be described as a channel that connects a \gls{Tx} and \pgls{Rx}. In order to support the typical sensing scenarios in a mobile radio network, both the network nodes and the \glspl{UE} may serve as the \gls{Tx} or the \gls{Rx} for sensing purpose. One of the goals of \gls{ISAC} is to accommodate both functionalities using one system and to sense both passive and active objects. Each entity in the integrated system may adjust its role according to the operating scenario. Fig.~\ref{fig:ISAC_operating_scenarios} gives an overview of eight exemplary operating scenarios and their applications in an \gls{ISAC} network.

			It is worth mentioning that modern mobile networks provide location services, although these services have only been introduced for commercial usage recently. Considering the \gls{UE} position as a sensing parameter of interest, the \gls{5G} mobile radio system providing location service is one form of \gls{ISAC} systems. Being the target objects themselves, these \glspl{UE} are actively involved in the radio transmission, with their positions being derived based on the radio link measurements obtained in the \gls{DL}, \gls{UL} or \gls{SL}~\cite{TS38.305}. This can be seen as active sensing, since the signal is received or transmitted by the object (\gls{UE}) in the sensing system. 
			
	\section{Integration}
		In this section, we identify three integration levels for a system with both sensing and communications capability, namely,
		\begin{itemize}
			\item Application level integration;
			\item Spectrum level integration;
			\item Full integration.
		\end{itemize}
		We describe the characteristics of the three approaches and further examine in-depth details of the fully integrated approach.
		
		\subsection{Application Level Integration}
			The first and straightforward form of integration involves integration of a sensing system and a communications system only at the application layer. Utilizing completely different hardware blocks and frequency bands, the two systems interact with each other only at the application layer in order to assist each other's functionality. An example of this, which has been intensively investigated, is \emph{sensor-assisted communications} whereby the results of the sensing system can assist the communications system to assign specific beam directions or communications resources \cite{AGN+20}.
			
		\subsection{Spectrum Level Integration}
			The second form of integration is referred to as spectrum level integration. It allows the spectrum to be shared on a time, frequency or spatial basis for the sensing system and for the communications system. This integration level also enables partial sharing of the hardware for the two systems, leading to reduced cost, area and weight. Furthermore, the shared resource (i.e. frequency, time or space) for the two systems could be dynamically split depending upon the specific requirements for each system.
			
		\subsection{Full Integration}
			The third form, namely full integration, enables sensing and communications to be simultaneously supported by one system using the same transmitted waveform and spectral resources. In this way, latency, cost, space and power consumption are optimally reduced. This integration level offers the most benefits, but also faces most challenges. To enable this form of integration, there are three key options: 1) add communications functionality to a sensing system, 2) add sensing functionality to a communications system or 3) investigate a new waveform optimized for both sensing and communications. We describe below the technical challenges of each of these options and review the potential way forward.
			
			\subsubsection{Add communications functionality to a sensing system}
\begin{figure}
					\centering
					\includegraphics[scale=0.55]{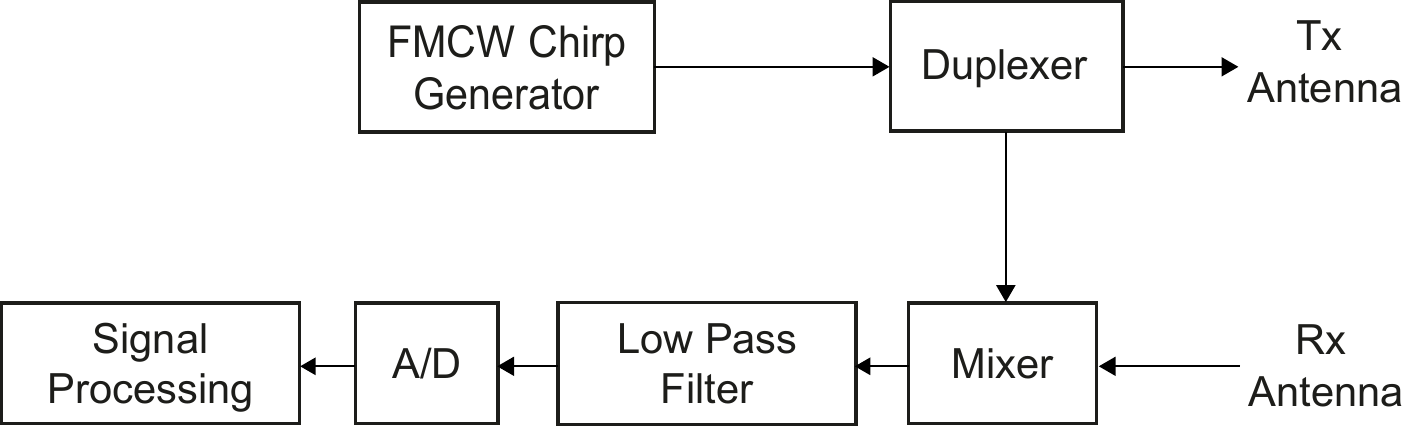}
					\sbox\glsscratchboxa{\footnotesize \gls{FMCW}}
					\caption{\unhcopy\glsscratchboxa~system - block diagram.}
					\vspace{-0.5cm}
					\label{fig:FMCW_block_diagram}
				\end{figure}

				The most popular waveform for sensing, as typically used for automotive radars, is \gls{FMCW}. \Pgls{FMCW} radar transmits a signal called a "chirp" which is a sinusoid whose frequency increases linearly with time. Specifically, an individual chirp $s_T(t)$  is characterized by the carrier frequency $f_{\text{c}}$, the bandwidth $B$ and the duration $T$, where the chirp slope is defined as $\alpha = B/T$,
				\begin{IEEEeqnarray}{rCl}
					s_T(t) &=& \exp\Big(\jj 2 \pi \big(f_{\text{c}} + 0.5\alpha t^2\big)\Big)
				\end{IEEEeqnarray}
				A block diagram of an exemplary \gls{FMCW} system is shown in Fig.~\ref{fig:FMCW_block_diagram}. 
				A series of \gls{FMCW} chirps is fed to the \gls{Tx} antenna and the mixer at the receiver. After mixing the transmitted signal and the received signals reflected from the target, the beat signal is generated, which has a frequency proportional to the delay corresponding to the distance between the radar and the target. The beat signal is passed through a low pass filter and \gls{ADC} for further signal processing implementation, e.g., target detection or sensing parameter estimation~\cite{AKC+20}.
				The main attractive feature of 
				\gls{FMCW} is its simplified hardware components and low-cost receiver. This is attributed to the low sampling-rate \gls{ADC}, due to the beat frequency, and to the cheap power amplifier, due to the low \gls{PAPR} of \gls{FMCW}. However, \gls{FMCW} has no modulation to convey communications data, rendering it difficult to be used for \gls{ISAC} applications. 
				
				A modification of \gls{FMCW} which has been considered for \gls{ISAC} is \gls{PC-FMCW} which combines phase coding with traditional linear frequency modulated waveform~\cite{Uys20}. 
				Another alternative for communications-enabled \gls{FMCW} is \gls{CPM-LFM} waveform, in which continuous phase modulation is combined with \gls{FMCW}~\cite{MSB+19}.
				Both \gls{PC-FMCW} and \gls{CPM-LFM} are capable of communications with low to medium data rate, e.g. 8 Mbps with a bandwidth of 500MHz. However, the capability of data transmission for 
				\gls{FMCW}-based waveforms is achieved at the expense of certain drawbacks compared to normal \gls{FMCW}, such as increased out-of-band emissions, high implementation complexity with high-rate \glspl{ADC}, 
				or degradation of sensing performance~\cite{MSB+19}.
				
				A further type of modified waveform based on \gls{FMCW} is partial chirp modulation which has been developed in~\cite{ANW+19}. It modulates every chirp only partially during the chirp cycle. A further variation is to embed the communications symbols inside different chirp cycles using time-frequency shift modulation~\cite{AOG+20}, in which different fractions of chirp are modulated with the time shift  related to the modulated symbol. Both schemes have been shown to have almost no deterioration of sensing performance and no increase in out of band emissions, but can only support low data rates.
				

			\subsubsection{Adding Sensing Functionality to a Communications System}
				In this section, we concentrate on adding sensing capability to the \gls{OFDM} waveform since \gls{OFDM} has become the dominating waveform for wireless communications systems, (i.e. LTE, \gls{5G} New Radio, IEEE Wi-Fi, etc.) in the last decade, given its robustness against multipath fading and simple receiver design. This multicarrier waveform facilitates adaptive transmission schemes and multiple access design across the subcarriers, offering flexibility in radio resource management. The investigation on \gls{OFDM} for joint data communications and radar sensing dates back to a decade ago~\cite{SW11}. Sensing parameters such as range and velocity can be estimated using the received symbols after the \gls{DFT}. When integrated with a communications system, these received symbols may be either pre-defined reference signals or demodulated data blocks. It has been shown that with appropriate parameterization, \gls{OFDM} can achieve the same sensing performance as chirp-based waveforms in terms of delay and Doppler estimation~\cite{FJ15}. Furthermore, estimation techniques developed for conventional sensing-only applications such as \gls{2D} \gls{DFT}, subspace techniques and compressed sensing can be easily exploited. With the present \gls{5G} \gls{NR} mobile communications system, a flexible air interface with configurable numerology has been developed and standardized. This potentially allows for some \gls{OFDM} radar sensing utilizing the same hardware while offering high data rate out-of-the-box. 
				
\begin{figure}
					\centering
					\includegraphics[scale=1]{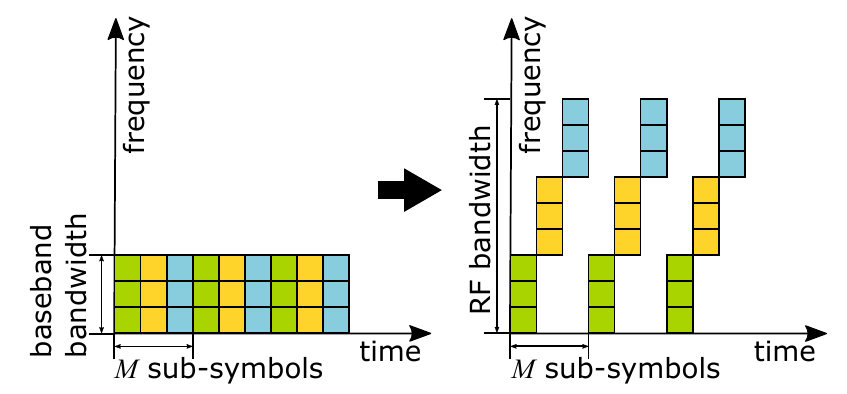}
					\vspace{-0.3cm}
					\sbox\glsscratchboxa{\footnotesize \gls{RF}}
					\caption{Narrow sub-symbols in baseband up-converted to stepped carrier frequency in \unhcopy\glsscratchboxa~band.}
					\vspace{-0.4cm}
					\label{fig:stepped_carrier_OFDM}
				\end{figure}

				One of the issues of \gls{OFDM} radar sensing, however, is that the processing bandwidth required for range estimation is much larger than that for a typical communications link if a fine resolution is desired. For example, a $\SI{15}{\centi\meter}$ range resolution requires $\SI{1}{\giga\hertz}$ bandwidth, which needs a high sampling rate \gls{ADC} with correspondingly high cost and high power consumption. Alternative solutions have been developed for generating an overall large bandwidth \gls{RF} signal using a narrow \gls{OFDM} baseband signal. In~\cite{SKS+18}, an \gls{OFDM} radar processing scheme with stepped carriers is proposed. The idea is to split up each wideband \gls{OFDM} symbol into $M$ sub-symbols. Each sub-symbol consists of a narrow sub-band transmitted sequentially in the baseband. This signal is up-converted to a stepped carrier frequency, generating a large \gls{RF} bandwidth as shown in Fig.~\ref{fig:stepped_carrier_OFDM}. 
				On the receiver side, the $M$ sub-symbols are combined to obtain the desired high resolution range profile corresponding to the large \gls{RF} bandwidth. The so-called stepped \gls{OFDM} scheme achieves high resolution using only a fraction of the equivalent channel bandwidth in the baseband; however, since the Doppler sampling rate is reduced by the factor $M$, the unambiguous velocity is reduced correspondingly. A similar scheme is presented in~\cite{KSS+18}, namely sparse \gls{OFDM}. Inspired by the frequency hopping concept, discrete but randomly chosen frequencies are used for up-conversion of the sub-symbols. This scheme preserves the resolution and unambiguity of a wideband \gls{OFDM} signal and enables compressed sensing processing to reconstruct the sub-Nyquist sampled signal. However, these advanced signal processing techniques may cause significant increase of computational complexity. Another alternative scheme in~\cite{NMM+20} shows that high range resolution and maximum unambiguous velocity can be maintained at the expense of higher \gls{RF} complexity. The idea is to multiply a narrow band \gls{OFDM} signal with a frequency comb consisting of $M$ carriers. Subsampling techniques are applied at the receiver side ensuring that the folded sub-bands do not overlap. Although all of these alternative solutions provide trade-offs between an improved range resolution and other limitations, they offer design flexibility for adapting the \gls{OFDM} system according to the various requirements of sensing applications.
				
				Another issue for an \gls{OFDM} sensing system is that it must support full-duplex when operating in monostatic scenarios. In order to enable long distance sensing with high dynamic range, stringent \gls{Tx}-\gls{Rx} isolation is required in a full-duplex transceiver. Due to the high \gls{PAPR} of \gls{OFDM}, providing good sensing performance in monostatic scenarios is still challenging for a compact \gls{UE}.
				
			\subsubsection{New waveform for both sensing and communications}
				Apart from the variations of \gls{FMCW} and \gls{OFDM} described above, tailored to accommodate \gls{ISAC}, alternative waveforms could be considered as potential candidates for providing an improved integration of the communications and sensing services especially in monostatic sensing scenarios. In this section, we review the state of the art of three such waveforms, namely \gls{OTFS}, \gls{OCDM} and \gls{AFDM} and analyze the benefits and drawbacks of their potential deployment. The choice of these particular waveforms was motivated not only by their ability to overcome shortcomings of \gls{OFDM} and \gls{FMCW} in the sensing and/or communications domain, but also by their compatibility with \gls{OFDM}. This is an important aspect for the industry, as it enables hardware reuse and backward compatibility of next generation wireless systems with their \gls{OFDM}-based predecessors, allowing for dynamic switching between them.
				
\begin{figure}
						\centering
						\includegraphics[scale=1.4]{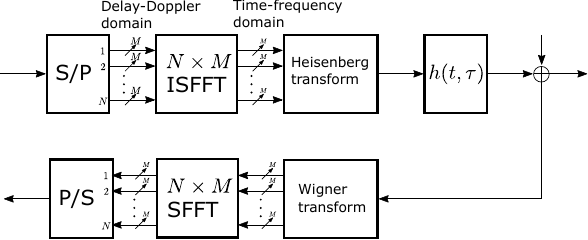}
						\sbox\glsscratchboxa{\footnotesize \gls{OTFS}}
						\caption{Block diagram of \unhcopy\glsscratchboxa~transmitter and receiver.}
						\label{fig:OTFS_block_diagram}
						\vspace{-0.3cm}
					\end{figure}

				\paragraph{\gls{OTFS}}
					\gls{OTFS} is a modulation scheme that multiplexes information symbols in the delay-Doppler domain. It can be viewed as a generalization of \gls{DFT}-spread-\gls{OFDM}, which spreads the information symbols both in the frequency and in the time domain. A \gls{2D} inverse \gls{SFFT}, also called \gls{OTFS} transform, is first applied to transform the symbols from the delay-Doppler to the time-frequency domain. This is followed by a Heisenberg transform, which is common for all multicarrier waveforms and for \gls{OFDM} is implemented by the \gls{IDFT} to obtain the time-domain transmit signal. The inverse process is applied at the receiver (Wigner transform and \gls{SFFT}). The block diagram of the \gls{OTFS} transmitter and receiver is shown in Fig.~\ref{fig:OTFS_block_diagram}. 
					The \gls{2D} intersymbol interference in the delay-Doppler domain demands for equalizers with higher computational complexity than \gls{OFDM}. Hence, a significant amount of research has been dedicated to the development of low-complexity detection schemes~\cite{GKC+20, SC20}.
					\gls{OTFS} is known for its ability to deal with high Doppler, as the time-variant channel appears constant in the delay-Doppler domain, as long as the velocity of the channel-related objects is constant. Hence, less frequent adaptation of the modulation and coding scheme is required; this can be particularly important for fast-varying channels, where even obtaining up-to-date channel state information at the transmitter can be challenging. Additionally, channel estimation and sensing are very closely related in \gls{OTFS}, as it operates in the delay-Doppler domain, where radar targets are localized, and it can inherently exploit sparsity in the delay-Doppler domain~\cite{SDH+19}. Furthermore, it can support \gls{MIMO} and offer a lower \gls{PAPR} than \gls{OFDM}, but its advantage diminishes with an increasing number of symbols in the \gls{OTFS} frame~\cite{SAC19}. As far as full-duplex and \gls{ADC} bandwidth are concerned, \gls{OTFS} faces the same challenges as \gls{OFDM}. Its sensing performance has been shown to be as accurate as that of \gls{OFDM} and \gls{FMCW}~\cite{GKC+20}. Additionally, delay-Doppler domain-based multiplexing of data and pilot symbols for channel estimation or multiplexing multiple users in the \gls{UL} might require the introduction of guard intervals~\cite{SDH+19, Had18}. 
					
\begin{figure}
						\centering
						\includegraphics[scale=1.2]{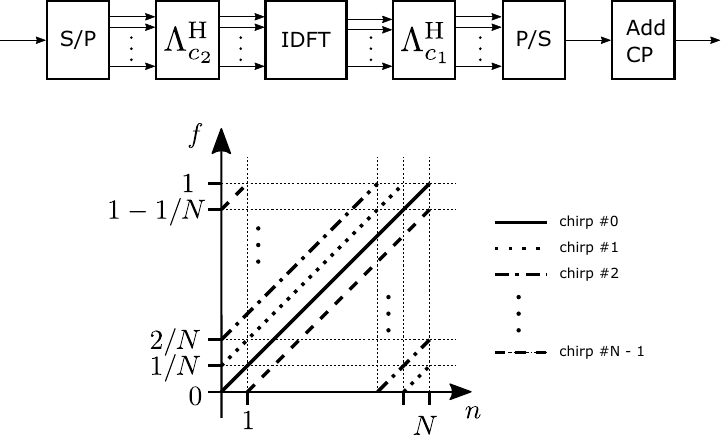}
						\vspace{-0.5cm}
						\sbox\glsscratchboxa{\footnotesize \gls{OCDM}}
						\sbox\glsscratchboxb{\footnotesize \gls{AFDM}}
						\caption{Block diagram of \unhcopy\glsscratchboxa/\unhcopy\glsscratchboxb~transmitter and resulting chirps in digital time-frequency domain.}
						\label{fig:OCDM_block_diagram}
						\vspace{-0.5cm}
					\end{figure}

				\paragraph{\gls{OCDM}}
					\gls{OCDM} is a multicarrier waveform that multiplexes a set of orthogonal chirps in time and frequency. Its implementation is based on the \gls{DFnT}. The \gls{DFnT} can be efficiently implemented with the use of the \gls{DFT}. The block digram of the \gls{OCDM} transmitter and a depiction of the resulting chirps are shown in Fig.~\ref{fig:OCDM_block_diagram}. 
					$\Lambda_{c_1}$ and $\Lambda_{c_2}$ are diagonal matrices with entries $[\Lambda_{c_i}]_{n,n} = \exp(\jj 2 \pi c_i n^2),\;c_i = 1/(2N),\; i=1,2,\;n=0,\ldots,N-1$. The chirps are shown with a different line style, with their wrapping stemming from their generation in the digital domain.
					Benefiting from the fact that the information-carrying chirps span the entire bandwidth, OCDM can extract full diversity in frequency selective channels and has been shown to outperform OFDM in frequency selective channels when a non-linear equalizer is employed~\cite{BZM+18}. An important advantage of chirp-based waveforms is that, similar to \gls{FMCW}, they can potentially support full-duplex operation, if the received reflected signal arrives before subsequent chirps are transmitted.  However, unlike \gls{FMCW}, the transmission and baseband bandwidths are the same (like \gls{OFDM}), leading to the requirement of high-rate \glspl{ADC}. A disadvantage of chirp-based waveforms compared to \gls{OFDM}, is that, with each chirp occupying the whole bandwidth, guard bands at the spectrum edges cannot be introduced. Additional operations are required to reduce out of band emissions~\cite{OM20}. Techniques like \gls{MIMO}, multi-user multiplexing and channel estimation (especially with scattered pilots), which are naturally supported by \gls{OFDM}, are not as straightforward with \gls{OCDM}. Although some works in this direction exist~\cite{OGN+21, OM19, WSH+21}, further research effort in these directions is required. The PAPR of OCDM has been shown to be identical to that of OFDM, while its sensing performance is expected to be slightly worse due to increased sidelobe level~\cite{OAN+20}.
				
				\paragraph{\gls{AFDM}}
					\gls{AFDM}, which was recently proposed by~\cite{BKK21}, is based on a discretization of the generalized Fresnel transform and has an almost identical structure to OCDM.  Their difference lies in the fact that, in contrast to \gls{OCDM}, where chirps have a fixed slope, \gls{AFDM} allows for adaptation of the chirp slope to the channel profile by optimizing $c_1$ and $c_2$. This enables \gls{AFDM} to extract full diversity of linear time-variant channels, outperforming \gls{OCDM} and having the same performance as \gls{OTFS}, but with lower implementation complexity~\cite{BKK21}. Due to its similarity with \gls{OCDM}, it is straightforward to conclude that \gls{AFDM} has the same characteristics in terms of \gls{PAPR}, full-duplex and \gls{ADC} bandwidth, as well as the same necessity for research in \gls{MIMO}, channel estimation and multi-user multiplexing aspects. No results on the sensing performance of \gls{AFDM} have been reported to date, but it is expected to be similar to that of \gls{OCDM}.
					
\begin{figure}
						\centering
						\includegraphics[scale=1]{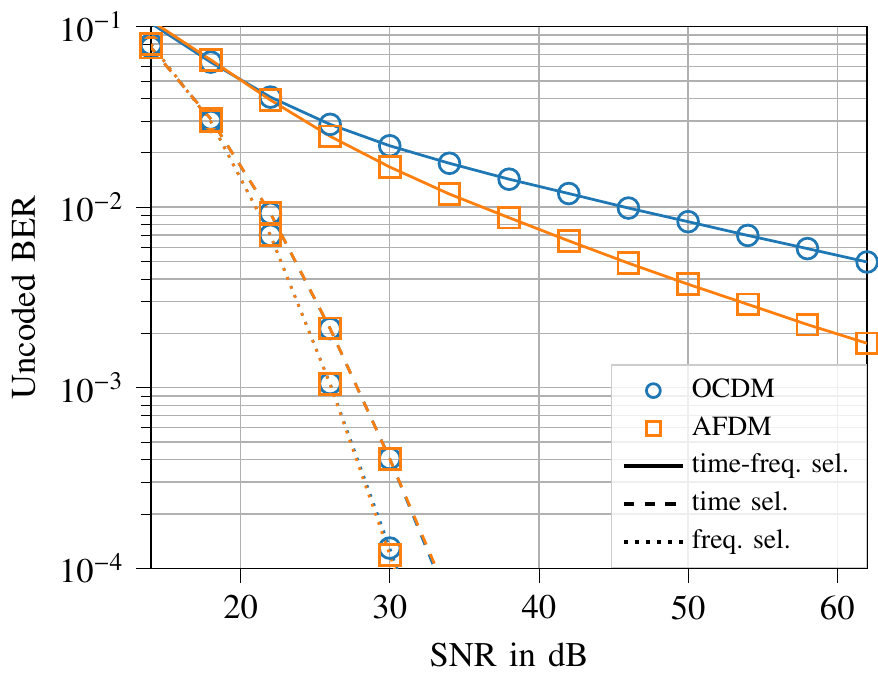}
						\sbox\glsscratchboxa{\footnotesize \gls{OCDM}}
						\sbox\glsscratchboxb{\footnotesize \gls{AFDM}}
						\caption{Comparison of \unhcopy\glsscratchboxa~and \unhcopy\glsscratchboxb~under frequency, time and time-frequency selective channels.}
						\label{fig:OCDM_AFDM_uncoded_BER_vs_SNR}
						\vspace{-0.3cm}
					\end{figure}

					To demonstrate the benefit of slope adaptation of \gls{AFDM}, in Fig.~\ref{fig:OCDM_AFDM_uncoded_BER_vs_SNR} we plot the uncoded \glspl{BER} vs. \gls{SNR} performance of \gls{OCDM} and \gls{AFDM} for frequency selective, time selective and time-frequency selective channels. 
					Following~\cite{BKK21}, we consider the channel to have paths with delays that are integer multiples of the sampling period and Doppler shifts that are integer multiples of the subcarrier spacing. The different paths have \gls{iid} Rayleigh fading with variance $1/P$, where $P$ is the number of paths. With $l_{\text{max}}$ denoting the maximum delay and $a_{\text{max}}$ denoting the maximum (absolute) Doppler shift, we consider three channels: a frequency selective channel with $l_{\text{max}}=2$ (3 paths in total), a time selective channel with $a_{\text{max}}=3$ (7 paths in total, as the Doppler shifts take values in $[-a_{\text{max}}, a_{\text{max}}]$) and a time-frequency selective channel with $l_{\text{max}}=2$ and $a_{\text{max}}=3$ (21 paths in total). The number of chirps is $N=64$, the number of symbols is $M=4$, 16-QAM symbols are transmitted and a linear \gls{MMSE} equalizer is used. For the time selective channel and the frequency selective channel, \gls{OCDM} and \gls{AFDM} exhibit the same performance. However, for the third channel, which is time and frequency selective, \gls{AFDM} clearly outperforms \gls{OCDM}, which, in contrast to \gls{AFDM}, fails to extract full multipath diversity.
					
	\section{Opportunities and Open Issues}
		The increasing interest in \gls{ISAC} has re-opened a large research area, not only because challenges for designing an integrated system are identified, but also because the topic connects two research communities which were relatively isolated from each other in the past. One of the major challenges is to bring the knowledge and experiences from both communities to work on the issues with joint efforts. Here, we provide examples of underlying challenges which deserve further study.
		\begin{enumerate}[label=(\alph*)]
			\item \textit{Air interface design aspects}: The integrated system requires a flexible air interface which offers the potential to meet diverse requirements of both functionalities simultaneously. This includes waveform design that has been considered in this work, suitable transmission schemes, antenna arrays, multi-beam optimization, multiplexing techniques and interference management.
			\item \textit{System design aspects}: The integration of the two functionalities raises new challenges to the system design, ranging from transceiver architecture to hardware engineering. For instance, full-duplex architectures commonly used for monostatic sensing devices have been for a long time a challenge for communications systems; a full-digital design found in most mobile radio terminals is a rather new concept for sensing devices. With respect to \gls{ISAC} applications for commercial use cases, requirements on implementation complexity, power consumption and cost potentially become more stringent. 
			\item \textit{Standardization aspects}: Mobile communications networks have shown that standardization efforts are essential for efficient spectrum utilization, interoperability of devices and guaranteed quality of service. The new applications enabled by the sensing capability integrated with a communications network may significantly change the resource management strategy. Given the gap between limited resources and limitless needs, decisions must be made on how to allocate resources efficiently.
		\end{enumerate}
	
	\section{Conclusion}
		In this work, we summarized key operating scenarios for \gls{ISAC} systems and identified three different levels for integrating the two functionalities in one system. For the full integration approach, we focused on the three design options and elaborated them in detail.
		
		Based on the observations presented and given the almost seamless deployment of mobile radio networks, adding sensing functionality to a communications system shows clear advantages. In addition, waveforms offering enhanced sensing capability while maintaining compatibility to \gls{OFDM}, are potentially interesting and worthy of further research.
		
		As presented, adding communications data to a sensing system (i.e., using \gls{FMCW}) can enable communications capability with limited data rate. Therefore, for systems which have a dedicated spectrum for sensing, e.g., the $\SI{77}{\giga\hertz}$ radar band, these schemes provide an opportunity for enabling \gls{ISAC}.
					
	\bibliographystyle{IEEEtran}
	\bibliography{IEEEabrv,references,MyBSTcontrol}

\end{document}